\documentclass[twoside,a4paper,12pt]{article}
\usepackage{color}
\usepackage{amsfonts}
\usepackage[psamsfonts]{amssymb}
\usepackage[bottom]{footmisc}
\usepackage{amsmath}
\usepackage{multicol}
\usepackage{a4wide}
\usepackage{fancyhdr}

\newcounter{eg}                         \newtheorem{eg}{Example}[section]
\def\beg{\begin{eg}\rm}                 \def\eeg{\hfill\sq\end{eg}}

\newcommand{\initiate}{\setcounter{equation}{0}} 
\def\c#1{{\cal #1}}

\def\Dirac{{\raise0.09em\hbox{/}}\kern-0.69em D}

\def\exterior{{{\raise0.2em\hbox{$\scriptstyle\bigwedge$}}{}}}

\def\kbar{{\mathchar'26\mkern-9muk}}

\def\lesssim{\mathrel{\hbox{\rlap{hbox{\lower8pt\hbox{$\sim$}}}\hbox{$<$}}}}

\def\sq{\hbox{\rlap{$\sqcap$}$\sqcup$}}

\def\p{\partial}

\def\dfrac #1#2{\displaystyle{\frac{#1}{#2}}}

\allowdisplaybreaks

\def\k{\kern-.1em\mathbin{,}\kern-.1em}
 
\def\hk{\kern.12em\raise-1em\hbox{$\hat{\raise1em\hbox{,}}$}\kern.12em}

\begin{document}

\title{Fuzzy de~Sitter  Space}

\author{Maja Buri\'c, Du\v sko Latas
                   and
       Luka Nenadovi\' c\thanks{majab@ipb.ac.rs, latas@ipb.ac.rs, lnenadovic@ipb.ac.rs}
                   \\[15pt]
        University of Belgrade,  Faculty of Physics, P.O. Box 44
                   \\
        SR-11001 Belgrade
       }

\date{}
\maketitle
\vfill
\parindent 0pt

\begin{abstract}
 We discuss properties of fuzzy de~Sitter space defined by means of
 algebra of the de~Sitter group $\mathrm{SO}(1,4)$ in  unitary irreducible 
representations. It was shown before that this fuzzy space has  
local frames with metrics that reduce, in the commutative limit,
to the de~Sitter metric. Here we determine  spectra of the embedding 
coordinates for  $(\rho,s=\frac 12)$ unitary irreducible
representations of the principal continuous series of the $\mathrm{SO}(1,4)$.
The result is obtained in the Hilbert space representation,
but using  representation theory it can be generalized to all 
representations of the principal continuous series.

\end{abstract}

\vfill
\setlength{\parskip}{10pt}

\pagestyle{plain}

\initiate
\section{Introduction}

Understanding of the structure of spacetime at very small scales is one of the most challenging problems in theoretical physics: more so as it is, as we commonly believe, related  to the properties of gravity at small scales, that is to  quantization of gravity. In the absence of a sufficient amount of experimental data, it is presently approached by mathematical methods: still there are basic tests which every model of quantum spacetime has to satisfy, as the mathematical consistency and the existence of a classical limit, usually to general relativity.

A feature  very often discussed in relation to quantization is discreteness of spacetime.  Discreteness can mathematically be imple\-men\-ted in various ways, for example by endowing  spacetime with  lattice or simplicial structure. When discreteness is introduced by means of representation of the position vector by noncommuting operators or matrices we speak of fuzzy spaces. Assumption that coordinates are operators comes from quantum mechanics: fact it is in quite natural (perhaps even too elementary) to presume that generalization of $\,[x^\mu,x^\nu] =0\,$ to $\,[x^\mu,x^\nu] \neq 0\,$ describes the shift of physical description to lower length scales. Operator representation has a potential  to solve various problems of classical gravity and quantum field theory: it introduces minimal length, which in the dual, momentum space, can in principle resolve the problem of UV divergences; singular configurations of gravitational field can potentially be dismissed as corresponding to non-normalizable states, and so on. In addition,  algebraic representation allows for a straightforward description of spacetime symmetries. Perhaps the main drawback of the assumption of discreteness is a loss of geometric intuition which is in many ways inbuilt in our understanding of  gravity.

 There are various ways to generalize geometry: one of the most important parts of any generalization is the definition of smoothness. In noncommutative geometry, derivatives are usually given by commutators; once they are defined, one can proceed more or less straightforwardly to differential geometry. We shall in the following use a variant of noncommutative differential geometry which was introduced by Madore,  known as the noncommutative  frame formalism, \cite{book}. It is a noncommutative generalization of the Cartan moving frame formalism and  gives a very  natural way to describe gravity on curved noncommutative spacetimes. In particular classical, that is commutative, limit of such noncommutative geometry is usually straightforward.

Let us introduce the notation. Noncommutative  space is an algebra $\c{A}$ generated by coordinates $x^\mu$ which are hermitian operators; fields are functions $\phi(x^\mu)$ on  $\c{A}$. Derivations or vector fields are represented by commutators. A special set of  derivations $e_\alpha$ can be chosen to define the moving frame,
\begin{equation}
 e_\alpha \phi =[p_\alpha, \phi], \qquad \phi\in\c{A}\, .
\end{equation}
 Derivations $e_\alpha$ are generated by antihermitian operators, momenta $p_\alpha$, which can but need not belong to algebra $\c{A}$.  1-forms $\theta^\alpha$ dual to $e_\alpha$ define the differential,
\begin{equation}
 \theta^\alpha(e_\beta) = \delta^\alpha_\beta,
\qquad d\phi = (e_\alpha \phi)\theta^\alpha.
\end{equation}
Supplementary condition which allows to interpret $\theta^\alpha$ as a locally orthonormal basis is $[\phi,\theta^\alpha]=0$. In addition, one imposes consistency constraints on both structures, algebraic (associativity) and differential ($d^2=0$), and  compatibility relations between them. 

General features of the noncommutative frame formalism and many applications to gravity are known, \cite{Madore:1997dt};
 the aim of our present investigation is to construct four-dimensional noncommutative spacetimes which correspond to known classical configurations of gravitational field. This means, to find algebras and  differential structures  which are noncommutative versions of, for example, black holes or  cosmologies. One very important idea in this context is that spacetimes of high symmetry can be naturally represented within the algebras of the symmetry groups. The first model of such noncommutative geometry was the fuzzy sphere \cite{fs}: it has a number of remarkable properties which make it a role example for understanding what  fuzzy geometry should or could mean. Different properties of the fuzzy sphere were used as guidelines to define other fuzzy spaces \cite{fuzzy},
including for us very important noncommutative de Sitter space in two and four dimensions \cite{Gazeau:2006hj,Gazeau:2009mi,Jurman:2013ota}. 
In our previous paper \cite{Buric:2015wta} we  analyzed differential-geometric properties of  fuzzy de Sitter space in four dimensions realized within  the algebra of the $\mathrm{SO}(1,4)$ group. We found two different differential structures with the de Sitter metric as  commutative limit. Here we analyze  geometry of fuzzy de~Sitter space that is the spectra of the embedding coordinates.

The plan of the paper is the following. In Section~2 we introduce notation for the $\mathrm{SO}(1,4)$, review some results of \cite{Buric:2015wta} and discuss the flat limit of fuzzy de~Sitter space revealing its relation to the Snyder space. In Section~3 we solve the eigenvalueproblem of coordinates  in the unitary irreducible representation $(\rho, s=\frac 12)\,$ of the principal continuous series. The obtaned spectrum we compare to the known group-theoretic result in Section~4.

\section{Metric and scaling limits}

We start with the algebra of the de~Sitter group $\mathrm{SO}(1,4)$ with generators $M_{\alpha\beta}$, ($\alpha, \beta = 0,1,2,3,4$) and  signature  $\,\eta_{\alpha\beta} = {\rm diag}(+----)$\,\footnote{Differently  from \cite{Buric:2015wta} we  here use the field-theoretic signature. Indices $\alpha$, $\beta$, \dots\ belong either to the set $\{0,1,2,3,4 \}$  or  $\{ 0,1,2,3\}$; in  cases when it is  not completely obvious we specify explicitly one the two sets. Indices  $i,j =1,2,3\dots$ \ are spatial.},
\begin{equation}
 [M_{\alpha\beta}, M_{\gamma\delta}] = - i(\eta_{\alpha\gamma} M_{\beta\delta}
 -\eta_{\alpha\delta} M_{\beta\gamma}-\eta_{\beta\gamma}
 M_{\alpha\delta}+  \eta_{\beta\delta} M_{\alpha\gamma}).
\end{equation}
The only $W$-symbol of the $\mathrm{SO}(1,4)$  group,  \cite{Herranz:1997us}, is the vector $W^\alpha$ which is quadratic in the generators
\begin{eqnarray}
& W^\alpha =\dfrac 18 \,\epsilon^{\alpha\beta\gamma\delta\eta}
 M_{\beta\gamma} M_{\delta\eta},                        \label{W}
\\[8pt]
& \ \ [M_{\alpha\beta}, W_\gamma]
= - i(\eta_{\alpha\gamma}W_\beta -\eta_{\beta\gamma}W_\alpha)           .       \label{**}\qquad
\end{eqnarray}
The Casimir operators of  are
\begin{equation}
 {\cal Q} = -\frac 12 \, M_{\alpha\beta} M^{\alpha\beta} ,
\qquad  \
{\cal W} =- W_\alpha W^\alpha  .
\end{equation}
The de Sitter algebra can be contracted to the Poincar\' e algebra by the In\" on\" u-Wigner contraction
\begin{equation}
M_{\alpha 4} \to \mu M_{\alpha 4},
\qquad M_{\alpha\beta}\to  M_{\alpha\beta}
\qquad   {\rm for}\ \  \mu \to \infty \,.                \label{inonu}
\end{equation}
 In the contraction limit $M_{\alpha 4}$  become  the generators of 4-translations while $M_{ij}$ and $M_{0i}$ generate 3-rotations and boosts. Further, $W_\alpha \to \mu W_\alpha\,$, $W_4\to W_4\,$ become  the components of the  Pauli-Lubanski vector of the Poincar\' e group (one can assume that  $W_4\to 0$). In the contraction limit $\,{\cal Q}\,$  and  $\,{\cal W}\,$ become the Casimir operators of the Poincar\' e group, ${\cal Q}\to \mu^2m^2$, ${\cal W}\to \mu^2 W^2$. Relations between the de Sitter and the Poincar\' e algebras exist also at the level of representations but not in general, only in some particular cases.

It is obvious that there is a strong analogy between commutative four-dimensional de~Sitter space described as an embedding in  five  flat dimensions,
\begin{equation}
 \eta_{\alpha\beta} {\tt x}^\alpha {\tt x}^\beta =-\,
 \frac 3\Lambda={\rm const},
\end{equation}
and the Casimir relation
\begin{equation}
 \eta_{\alpha\beta}W^\alpha W^\beta = -\c{W} ={\rm const}.
\end{equation}
It is therefore natural identify $W^\alpha$ with the embedding coordinates, as first proposed in \cite{Gazeau:2006hj},
\begin{equation}
 x^\alpha = \ell \, W^\alpha                   \label{coord}
\end{equation}
and to define fuzzy  de~Sitter space as a unitary irreducible representation (UIR) of the $\mathrm{so}(1,4)$ algebra. This definition makes sense\footnote{That is, it has a straightforward meaning; see a comment related to double scaling limits given below.} in all cases except when $\c{W}=0$, that is,  for  Class-I irreducible representations.

Group generators are dimensionless so a constant $\,\ell$ is introduced in (\ref{coord}) to give $x^\alpha$ a dimension of length\footnote{As we use units in which $\,\hbar=1$, momenta have dimension of the inverse length.}. There are two scales in our problem: the cosmological constant, $\Lambda\sim (10^{26}{\rm m})^{-2}$, and the Planck length, $\ell_{Pl}\sim 10^{-35} {\rm m}$. In preference to $ \ell_{Pl} $ 
we will use third constant, noncommutativity parameter  $\kbar$, and assume that
\begin{equation}
  \ell_{Pl}^2 < \kbar < (10^{-19}{\rm m})^2.   \label{estimate}
\end{equation}
The upper bound in (\ref{estimate}) is a rough experimental limit to $\kbar$.  Assumption which is often taken, $\kbar\approx\ell_{Pl}^2 $, would indicate that noncommutative geometry is directly related to quantization of gravity; a more moderate (\ref{estimate}) means that noncommutative geometry  is or might be an effective description of quantized gravity in the appropriate range of distances. We thus scale coordinates as
\begin{equation}
x^\alpha = \kbar \sqrt{\frac\Lambda 3} \,\, W^\alpha   .   \label{qq}
\end{equation}
Then the relation between the quartic Casimir operator and the cosmological constant reads
\begin{align}
{\cal W}= \frac{9}{\kbar^2\Lambda^2} \,.
\end{align}

Relations as (\ref{qq}) define the quantization condition. Dimensionally, we could have  assumed a more general relation of the form
\begin{equation}
 x^\alpha = c(\kbar\Lambda)^{-n} \,\kbar \sqrt\Lambda \, W^\alpha\, ,
\end{equation}
but  we chose the simplest, $n=0$.  For an interesting discussion of the quantization condition defined with respect to the Compton wavelength of the elementary system, see \cite{Gazeau:2009mi}.

Limit  $\kbar\to 0\,$ is the commutative limit of  fuzzy de Sitter space. From
\begin{equation}
 [W^\alpha, W^\beta] = -\frac i2\, \epsilon^{\alpha\beta\gamma\delta\eta}\,
W_\gamma M_{\delta\eta}                \label{[W,W]}
\end{equation}
we see that position commutator is proportional to $\kbar$,
\begin{equation}
 [x^\alpha, x^\beta] = -\frac i2\,\sqrt{\frac \Lambda 3}\,\,\kbar \epsilon^{\alpha\beta\gamma\delta\eta}
x_\gamma M_{\delta\eta} ,                 \label{[x,x]}
\end{equation}
that is, for $\kbar\to 0\,$ coordinates commute. The flat (noncommutative) limit on the other hand can be obtained when we consider de~Sitter space in a `small neighbourhood' of a specific point, for example at the north pole,
\begin{equation}
 x^4 \approx \sqrt{\frac 3\Lambda}\, , \qquad x^\alpha \approx 0 
\qquad (\alpha = 0,1,2,3)
\end{equation}
for $ \Lambda\to 0$. At the level of the symmetry group this limit is defined by the  In\" on\" u-Wigner contraction (\ref{inonu}). Commutation relations contract to
\begin{equation}
 [x^4,x^\alpha] =-\frac i2\,\sqrt{\frac \Lambda 3}\,\,
 \epsilon^{4\beta\gamma\delta\eta}\,
x_\gamma M_{\delta\eta} \to 0
\end{equation}
and it is consistent to take $\,  x^4 =\sqrt{\frac 3\Lambda}\,$= const.  Furthermore,
\begin{equation}
 [x^\alpha, x^\beta] = -\frac i2\,\kbar
\epsilon^{\alpha\beta\gamma\delta 4}
\left( \frac {1}{\mu^2} M_{\gamma \delta}
 +\sqrt{\frac \Lambda 3}\,x_\gamma
  M_{4\delta}\right)\to -\frac i2\,\frac{\kbar}{\mu^2}\,
\epsilon^{\alpha\beta\gamma\delta 4}
 M_{\gamma \delta}     \, .
\end{equation}
Denoting $ \, {\kbar}/({2\mu^2} )={ a^2}$, we see that we obtained the dual to the Snyder algebra. Namely, we found
\begin{equation}
 [x^i,x^j]\sim ia^2 \epsilon^{ijk} M_{0k},\qquad
[x^0,x^i]\sim ia^2 \epsilon^{ijk} M_{jk},
\end{equation}
whereas the position algebra of \cite{Snyder:1946qz} reads
\begin{equation}
 [x^i,x^j]\sim ia^2  M^{ij},\qquad
[x^0,x^i]\sim ia^2 M^{0i}.
\end{equation}
The limit $\mu\to\infty$ corresponds to $a\to 0$.

In  \cite{Buric:2015wta},   two sets of momenta that define fuzzy geometries with correct commutative limits to classical de~Sitter space  were proposed.
In the noncommutative frame formalism,  fulfil stricter requirements than coordinates: first, they  close into an algebra which is at most quadratic. In addition, if we wish to interpret tetrad $e^\alpha_A $ and  metric $ \, g^{\alpha\beta} =  \eta^{AB} e^\alpha_A \, e^\beta_B  \, $ as fields, we have to require that the frame elements depend only on coordinates,
\begin{equation}
 [p_A, x^\alpha]\equiv e^\alpha_A =e^\alpha_A(x), \qquad \ \ x\in{\cal A}.             \label{frameelements}
\end{equation}
It is simplest to choose $p_A$ among the group generators\footnote{See, however,  comments given in the Appendix.}. When momenta close into a Lie algebra, $\,  [p_A, p_B] = C^D{}_{AB}\, p_D\, $, the curvature defined in the framework of the noncommutative frame formalism is constant \cite{book}, and the curvature scalar is given by
\begin{equation}
 R = \frac 14\, C^{ABD} C_{DAB}.                        \label{curvature}
\end{equation}
This means in particular that, in our case, momenta scale as $ \sqrt{ \Lambda}\, $.

If we wish  to preserve the full de~Sitter symmetry on fuzzy de Sitter space, we choose as momenta  all ten generators $M_{\alpha\beta}$,
\begin{equation}
i p_A = \sqrt{ \zeta\Lambda}\, M_{\alpha\beta} ,
                                                   \label{pA}
\end{equation}
where index $A=1,\dots, 10 $, denotes antisymmetric pairs $[\alpha\beta]$. Normalization of the scalar curvature to $\, R=4\Lambda\,$ gives $\,\zeta = 1/3$. There are ten frame 1-forms $\theta^A$. Assuming the flat frame,   $g^{AB}=\eta^{AB}$,  with signature $\, (++++++----)$, for the spacetime components of the metric, $\,  g^{\alpha\beta} = e^\alpha_A e^\beta_B \eta^{AB} \ (\alpha = 0,1,2,3,4) $,
we obtain
\begin{equation}
 g^{\alpha\beta} =
\eta^{\alpha\beta} - \frac{ \Lambda}{3}\, x^\beta x^\alpha .
\end{equation}
In the commutative limit $\,g^{\alpha\beta}$ is singular and reduces to the projector on four-dimensional de~Sitter space.

The second choice of  momenta is
\begin{equation}
 i\tilde p_0 =\sqrt{\tilde\zeta\Lambda} \, M_{04},\qquad
 i\tilde p_i =\sqrt{\tilde\zeta\Lambda} \,(M_{i4}+M_{0i}),
 \quad i=1,2,3    .   \label{444}
\end{equation}
There are now four frame 1-forms $\,\tilde\theta^\alpha$, $\alpha = 0,1,2,3$. Calculating the spacetime components of the metric, for (the noncommutative
equivalent of) the line element we find
\begin{equation}
 \tilde ds^2 = (\tilde\theta^0)^2- (\tilde\theta^i)^2 =
\tilde d\tau^2 - e^{-\frac{2\tau}{\ell}} \,(\tilde dx^i)^2
\end{equation}
with identification of the cosmological time $\tau$
\begin{equation}
 \frac \tau\ell = -\log \left( \frac{x^0+x^4}{\ell}
 \right).                             \label{time}
\end{equation}
This noncommutative metric and the corresponding moving frame  do not possess the complete  de~Sitter symmetry. Normalization of the scalar curvature to the usual value gives $\,\tilde\zeta = 16/3$.

\section{Coordinates}

Let us consider the spectra of the embedding  coordinates $x^\alpha$. Classification of the unitary irreducible representations of the de Sitter group  was  done in \cite{U,I,R}; the UIR's of the $\mathrm{SO}(1,4)$ are induced from representations of its maximal compact subgroup $\mathrm{SO}(4)$. The representation  basis $\,\{ f^{k,k'}_{m,m'}\} $ is  discrete ($k$ and $k^\prime$   label the UIR's  of the two $\mathrm{SO}(3)$ subgroups of $\mathrm{SO}(4)$). The unitary irreducible of the $\mathrm{SO}(1,4)$ are infinite-dimensional, labelled by two quantum numbers, $\rho$ (or $\nu=i\rho$, $q=1/2+i\rho$) and $s$.\footnote{In comparison with \cite{R}, $p=s$, $\sigma = \frac 14+\rho^2$.} They are grouped in three series,
\begin{itemize}
  \item[--] principal continuous series,
$\rho\in {\bf R}$, $\rho\geq 0$,\  $s = 0, \frac 12, 1, \frac 32,\dots$\\
 ${\cal Q} = -s(s+1)+ \frac 94 + \rho^2$,\quad
 ${\cal W} =  s(s+1)( \frac 14 + \rho^2)$,
  \item[--]  complementary continuous series,
 $\nu\in {\bf R}$, \ $\vert\nu\vert<\frac 32  $,
\ $s = 0, 1, 2\dots$ \\
${\cal Q} = -s(s+1)+ \frac 94 - \nu^2$,\quad
 ${\cal W} =  s(s+1)( \frac 14 - \nu^2)$, and
  \item[--]  discrete series,
\ $s =  \frac 12, 1, \frac 32,2 \dots$,
\ $q= s,s-1,\dots 0\ {\rm or}\ \frac 12$ \\
${\cal Q} = -s(s+1) - (q+1)(q-2)$,\quad
 ${\cal W} = - s(s+1)q(q-1)$.
\end{itemize}
In the discrete case there are two inequivalent representations  $\pi^\pm_{s,q}\,$ for each value of $q$ and $s$; values of the Casimir operators are  discrete.

Using known matrix elements of  $M_{\alpha\beta}$ from \cite{R}, one can calculate  matrix elements of $W^\alpha$ in  basis   $\, \{f^{k,k'}_{m,m'} \}$. We find
\begin{align}
{W_0}\,f^{k,k'}_{m,m'}=&\left(k'(k'+1)-k(k+1)\right) f^{k,k'}_{m,m'}
\label{w_0}  \, ,\\[6pt]
{W_4}\, f^{k,k'}_{m,m'} =&
 -\dfrac{i}{2} A_{k,k'} (k-k')\left( \sqrt{(k-m+1)(k'+m'+1)}\, f^{k+\frac{1}{2},k'+\frac{1}{2}}_{m-\frac{1}{2},m'+\frac{1}{2}} \right.
\label{w_4} \\
& \phantom{-\dfrac{i}{2} A_{k,k'} (k-k')}
\left.- \sqrt{(k+m+1)(k'-m'+1)}\, f^{k+\frac{1}{2},k'+\frac{1}{2}}_{m+\frac{1}{2},m'-\frac{1}{2}} \right)
\nonumber \\
&-\frac{i}{2} B_{k,k'} (k+k'+1)\left( \sqrt{(k+m)(k'+m'+1)}\, f^{k-\frac{1}{2},k'+\frac{1}{2}}_{m-\frac{1}{2},m'+\frac{1}{2}}\right.
\nonumber \\
&\phantom{-\frac{i}{2} B_{k,k'} (k+k'+1)} \left. + \sqrt{(k-m)(k'-m'+1)}\, f^{k-\frac{1}{2},k'
+\frac{1}{2}}_{m+\frac{1}{2},m'-\frac{1}{2}} \right)
\nonumber \\
&-\frac{i}{2} C_{k,k'} (k+k'+1)\left( \sqrt{(k-m+1)(k'-m')}\, f^{k+\frac{1}{2},k'-\frac{1}{2}}_{m-\frac{1}{2},m'+\frac{1}{2}}
\right.   \nonumber \\
&\phantom{-\frac{i}{2} C_{k,k'} (k+k'+1)}
\left. + \sqrt{(k+m+1)(k'+m')}\, f^{k+\frac{1}{2},k'-\frac{1}{2}}_{m+\frac{1}{2},m'-\frac{1}{2}} \right)
\nonumber  \\
&-\frac{i}{2} D_{k,k'} (k-k')\left( \sqrt{(k+m)(k'-m')}\, f^{k-\frac{1}{2},k'-\frac{1}{2}}_{m-\frac{1}{2},m'+\frac{1}{2}}
\right.  \nonumber \\
&\phantom{-\frac{i}{2} D_{k,k'} (k-k')}
\left. - \sqrt{(k-m)(k'+m')}\, f^{k-\frac{1}{2},k'-\frac{1}{2}}_{m+\frac{1}{2},m'-\frac{1}{2}}\right) \, ,
 \nonumber \\[6pt]
{W_3}\, f^{k,k'}_{m,m'} =&\dfrac{1}{2} A_{k,k'}
\left( (m-k'+2m') \sqrt{(k-m+1)(k'+m'+1)}\, f^{k+\frac{1}{2},k'+\frac{1}{2}}_{m-\frac{1}{2},m'+\frac{1}{2}} \right.
 \\
&\phantom{ \dfrac{1}{2} A_{k,k'} }
\left. - (k+k'+2m') \sqrt{(k+m+1)(k'-m'+1)}\, f^{k+\frac{1}{2},k'+\frac{1}{2}}_{m+\frac{1}{2},m'-\frac{1}{2}} \right.
\nonumber \\
&\phantom{ \dfrac{1}{2} A_{k,k'} }
\left. -\sqrt{(k-m)(k+m+1)(k+m+2)(k'+m'+1)}\, f^{k+\frac{1}{2},k'+\frac{1}{2}}_{m+\frac{3}{2},m'+\frac{1}{2}} \right)
\nonumber \\
& + \dfrac{1}{2} B_{k,k'}
\left( -(m-k'+2m') \sqrt{(k+m+1)(k'+m'+1)}\, f^{k-\frac{1}{2},k'+\frac{1}{2}}_{m-\frac{1}{2},m'+\frac{1}{2}} \right.
\nonumber \\
& \phantom{ +\dfrac{1}{2} B_{k,k'} }
\left. + (k-k'-2m'+1) \sqrt{(k-m)(k'-m'+1)}\, f^{k-\frac{1}{2},k'+\frac{1}{2}}_{m+\frac{1}{2},m'-\frac{1}{2}} \right.
\nonumber \\
&\phantom{ +\dfrac{1}{2} B_{k,k'} }
\left. -\sqrt{(k-m-1)(k-m)(k+m+1)(k'+m'+1)}\, f^{k-\frac{1}{2},k'+\frac{1}{2}}_{m+\frac{3}{2},m'+\frac{1}{2}} \right)
\nonumber \\
&  +\dfrac{1}{2} C_{k,k'}
\left( (m+k'+2m'+1) \sqrt{(k-m+1)(k'-m')}\, f^{k+\frac{1}{2},k'-\frac{1}{2}}_{m-\frac{1}{2},m'+\frac{1}{2}} \right.
\nonumber \\
& \phantom{ +\dfrac{1}{2} C_{k,k'} }
\left. + (k-k'+2m'-1) \sqrt{(k+m+1)(k'+m')}\, f^{k+\frac{1}{2},k'-\frac{1}{2}}_{m+\frac{1}{2},m'-\frac{1}{2}} \right.
\nonumber \\
& \phantom{ +\dfrac{1}{2} C_{k,k'} }
\left. -\sqrt{(k-m)(k+m+1)(k+m+2)(k'-m')}\, f^{k+\frac{1}{2},k'-\frac{1}{2}}_{m+\frac{3}{2},m'+\frac{1}{2}} \right)
\nonumber \\
& +\dfrac{1}{2} D_{k,k'}
\left(- (m+k'+2m'+1) \sqrt{(k+m)(k'-m')}\, f^{k-\frac{1}{2},k'-\frac{1}{2}}_{m-\frac{1}{2},m'+\frac{1}{2}} \right.
\nonumber \\
&\phantom{ +\dfrac{1}{2} D_{k,k'} }
\left. - (k+k'-2m'+2) \sqrt{(k-m)(k'+m')}\, f^{k-\frac{1}{2},k'-\frac{1}{2}}_{m+\frac{1}{2},m'-\frac{1}{2}} \right.
\nonumber \\
&\phantom{ +\dfrac{1}{2} D_{k,k'} }
\left. -\sqrt{(k-m-1)(k-m)(k+m+1)(k'-m')}\, f^{k-\frac{1}{2},k'-\frac{1}{2}}_{m+\frac{3}{2},m'+\frac{1}{2}} \right)\, .
\nonumber
\end{align}

Constants $ A_{k,k'}$, $ B_{k,k'}$, $ C_{k,k'}$, $ D_{k,k'}$ are given for each concrete representation in \cite{R}. From (\ref{w_0}) we see  that $W_0$ has  discrete spectrum as noted in  \cite{Gazeau:2006hj}. On the other hand, the eigenvalue equation for $W_4$  (and likewise for $W_i$) is quite difficult, if at all possible, to solve in this basis.

We therefore restrict to simpler problem: to find the eigenvalues of $W^\alpha$ for a specific class of representations.
The simplest possibility would be to consider Class I UIR's  (they are in the  principal and complementary series): their Hilbert space representations are known, they have a lowest weight state so the coherent states can be constructed,  
etc. However, Class I is characterized  by condition $\,{\cal W}=0$: thus in our framework these UIR's cannot be simply interpreted as  de~Sitter spaces: a fixed $\kbar$ implies $\Lambda\to\infty$\footnote{It is on the other hand certainly possible to define specific double scaling limits, in order to interpret Class I representations as fuzzy de~Sitter spaces; this point remains to be explored.}. Another subset which is singled out  mathematically and physically is the principal continuous series. As  shown in \cite{Mickelsson:1972fh}, in the Wigner-In\"on\"u contraction limit these UIR's contract to a sum of two representations of the Poincar\' e group with positive value of the mass-squared. The Hilbert space representations of the principal continuos series were found in \cite{Moylan,Bohm:1985zn}: we shall perform the construction explicitly in the simplest nontrivial case,  $s=1/2$.

We start from the $s=0\,$ representation of the principal continuous series. The representation  space  is a direct sum of the two $s=0\,$ representation spaces of the Poincar\' e algebra, \cite{Moylan}. The states in each summand are  wave functions in momentum space $\psi(\vec p)$, with the scalar product given by
\begin{equation}
 (\psi,\psi^\prime )=\int \frac{d^3p}{2p_0}\, \psi^*\psi^\prime\, ,
\end{equation}
and $ p_0=\sqrt{- p_i p^i +m^2}$\, . Generators of the $\mathrm{SO}(1,4)$ group $\, M_{\alpha\beta}\vert_{s=0}\equiv L_{\alpha\beta}\, $ are
\begin{align}
L_{ij} =& i\left( p_i \, \frac{\partial}{\partial p^j }-
p_j \, \frac{\partial}{\partial p^i } \right) \label{11}\\
L_{0i} =& ip_0\, \frac{\partial}{\partial p^i }
 \\
L_{40} =& -\frac{\rho}{m}\, p_0 +\frac{1}{2m}\, \{p^i, L_{0i}\}\\
L_{4k} =& -\frac{\rho}{m}\, p_k -\frac{1}{2m}\, \{p^0, L_{0k}\}
-\frac{1}{2m}\, \{p^i, L_{ik}\}  .\label{44}
\end{align}
They are hermitian with respect to the given scalar product, and one can easily check that $\,W^\alpha\vert_{s=0} =0\,$, therefore $\,{\cal W}=0\,$ for $\,(\rho, s=0)$.

Higher spin representations $(\rho,s)$ can be obtained from $(\rho, s=0)\,$ by adding spin generators $\,S_{\alpha\beta}$  to  orbital generators $\,L_{\alpha\beta}$.  Representation space will be again a direct sum of two spaces,  each  equivalent to the Hilbert space of the Bargmann-Wigner representation of the Poincar\'e group of a fixed spin $s$, \cite{Bargmann:1948ck}. We shall here discuss the eigenvalue problem for ${s=\frac 12}\,$; the case of higher spins is more involved because of an additional projection to the highest spin states, \cite{BandaGuzman:2016mau}. In addition, we will consider just a `half' of the representation space, the other half being equivalent, \cite{Bohm:1985zn}.

States for ${s=\frac 12}\,$ are Dirac bispinors in  momentum space $\psi(\vec p)$ which are solutions to the Dirac equation. The  Bargmann-Wigner scalar product is given by 
\begin{equation}
 (\psi,\psi^\prime)=\int \frac{d^3p}{\vert p_0\vert}\,
 \psi^\dagger \gamma^0\psi^\prime
=\int \frac{d^3p}{ p_0^2}\, \psi^\dagger\psi^\prime\, .   \label{scalar}
\end{equation}
In the Dirac representation of  $\gamma$-matrices,\ $ \gamma^0= \begin{pmatrix}
     I & 0\\ 0 & -I
 \end{pmatrix}$, \ $\gamma^i=  \begin{pmatrix}
     0 & \sigma_i\\ -\sigma_i & 0
 \end{pmatrix} $,
the states are bispinors
\begin{equation}
 \psi(\vec p) =\begin{pmatrix}
     \varphi(\vec p)\\[4pt]
 -\,\dfrac{\vec p\cdot\vec \sigma}{p_0+m}\,\varphi(\vec p)
 \end{pmatrix} \,
\end{equation}
and the scalar product reduces to
\begin{equation}
 (\psi,\psi^\prime)=\int \frac{d^3p}{ p_0}\,
 \frac {2m}{p_0+m}\, \, \varphi^\dagger \varphi^\prime  \, .
\end{equation}
In the chiral representation which we will use later,
$ \,\tilde\gamma^0= \begin{pmatrix}
     0 & I\\ I & 0
 \end{pmatrix}$, \ $\tilde\gamma^i=  \begin{pmatrix}
     0 & -\sigma_i\\ \sigma_i & 0
 \end{pmatrix} \, $
and the states  can be parametrized as
\begin{equation}
\tilde  \psi(\vec p) =\begin{pmatrix}
   \tilde\chi(\vec p)\\[4pt]
 \dfrac{p_0+\vec p\cdot\vec \sigma}{m}\,\tilde\chi(\vec p)
 \end{pmatrix}                            \label{chiral}
\end{equation}
while the scalar product becomes
\begin{equation}
 (\psi,\psi^\prime)= (\tilde\psi,\tilde\psi^\prime)=
\int \frac{d^3p}{ p_0}\,\frac{2}{m}\,\,
 \tilde \chi^\dagger \, (p_0+\vec p\cdot\vec \sigma)
\,\tilde \chi^\prime  \, .                   \label{chiralproduct}
\end{equation}

The de~Sitter group generators  are given by
\begin{align}
M_{ij} =& L_{ij}+S_{ij},\, \quad S_{ij}= \frac i4\,
[\gamma_i,\gamma_j] ,\\
M_{0i} =& L_{0i}+S_{0i}, \quad  S_{0i}= \frac i4\,
[\gamma_0,\gamma_i] ,\\[4pt]
M_{40} =& -\frac{\rho}{m}\, p_0 +\frac{1}{2m}\, \{p^i, M_{0i}\} ,\\[4pt]
M_{4k} =& -\frac{\rho}{m}\, p_k -\frac{1}{2m}\, \{p^0, M_{0k}\}
-\frac{1}{2m}\, \{p^i, M_{ik}\}  .
\end{align}
One can easily check that with respect to (\ref{scalar}) all generators are hermitian:  for an operator-valued  $M$ of the 2$\times$2 block-form
\begin{equation}
 M=\begin{pmatrix}
     A & B\\
     B & A
 \end{pmatrix}
  \label{ABBA}
\end{equation}
hermiticity condition reads, in the Dirac representation of $\gamma$-matrices,
\begin{equation}
p_0^{-1}  A = A^\dagger  p_0^{-1}, \qquad
p_0^{-1}  B = - B^\dagger   p_0^{-1} .                   \label{abba}
  \end{equation}

From (\ref{11}-\ref{44}) we find  the components $W^\alpha$:
\begin{align}
{W^0} =& -\frac{1}{2m} \begin{pmatrix}
  (\rho - \frac i2)\, p_i\sigma^i
+i\, p_0^2\, \frac{\partial}{\partial p^i}\sigma^i
& \epsilon^{ijk}p_0p_i\, \frac{\partial}{\partial p^j}\, \sigma_k +
\frac{3i}{2}\, p_0 \\[8pt]
 \epsilon^{ijk}p_0p_i\,\frac{\partial}{\partial p^j}\, \sigma_k +
\frac{3i}{2}\, p_0
& (\rho - \frac i2)\, p_i\sigma^i +i p_0^2\,
 \frac{\partial}{\partial p^i}\, \sigma^i
 \end{pmatrix}
\\[8pt]
{ W^4} =& -\frac 12 \begin{pmatrix}
  i p_0\, \frac{\partial}{\partial p^i}\sigma^i
&  \epsilon^{ijk} p_i\, \frac{\partial}{\partial p^j}\, \sigma_k +
\frac{3i}{2} \\[4pt]
 \epsilon^{ijk} p_i\, \frac{\partial}{\partial p^j}\, \sigma_k +
\frac{3i}{2}
&  i p_0\, \frac{\partial}{\partial p^i}\sigma^i
 \end{pmatrix}\\[8pt]
{ W^i} =&  \begin{pmatrix}
U^i
& V^i \\[8pt]
 V^i & U^i
 \end{pmatrix}  ,
\end{align}
with
\begin{align*}
U^i =& \frac{p_0}{2m}\left( -i p^i\frac{\partial}{\partial p_k}\,\sigma_k
+(\rho -\frac i2 )\, \sigma^i \right)  , \\[4pt]
V^i=&\frac{1}{2m}\left( -2i p^i +i\epsilon^{ijk}p_l M_{jk}\sigma^l
+\epsilon^{ijk}\big(
i\rho p_j- p_lp^l \frac{\partial}{\partial p^j}
- p_jp_l \frac{\partial}{\partial p_l} \big)\sigma_k \right) .
\end{align*}

We have seen already in (\ref{w_0}) that the spectrum of $W^0\,$ is discrete in every UIR of the $\mathrm{SO}(1,4)$ group, and that the eigenvalues are $\, k'(k'+1)-k(k+1)\, $. On the other hand, due to de~Sitter symmetry, spatial directions $i$ and 4 are equivalent: therefore $W^4$ and $W^i$  have the same spectra. We can thus confine to the eigenvalue problem  of $W^4$.

We proceed as follows. First, we observe that in the Dirac representation $W^4$  has the form (\ref{ABBA}) with
\begin{equation}
 A= -\frac i2\, p_0\, \frac{\partial}{\partial p^i}\sigma^i ,
\qquad B=-\frac 12 \left( \epsilon^{ijk} p_i\,
\frac{\partial}{\partial p^j}\, \sigma_k +
\frac{3i}{2}\right)  .                       \label{AandB}
\end{equation}
Unitary transformation to the chiral representation transforms  $W^4$ to
\begin{equation}
\tilde W^4= UW^4U^\dagger = \begin{pmatrix}
                            A + B & 0\\
                            0 & A - B
                           \end{pmatrix} ,
\end{equation}
with $\, U =
\dfrac{1}{\sqrt 2}\begin{pmatrix}
                            I & I\\ I &-I
                           \end{pmatrix} \,$,
and we can solve the eigenvalue problems for $A+B$ and $A-B$ separately. But in fact,  one can easily check that if $\, \tilde \chi\,$ satisfies
\begin{equation}
 (A+B)\, \tilde\chi = \lambda\, \tilde \chi  \, ,     \label{353}
\end{equation}
the other component of the eigenvalue equation,
\begin{equation}
 (A-B)\, \tilde\chi = \lambda\,\,
 \dfrac{p_0+\vec p\cdot\vec \sigma}{m}\, \tilde \chi  \, ,
\end{equation}
is automatically satisfied for $A$, $B$ given by (\ref{AandB}). 

Since $W^4$ commutes with the generators  of 3-rotations, we can diagonalize $A+B\,$ simultaneously with $M_{ij}$, that is, we can write the eigenfunctions in the form
\begin{equation}
\tilde \chi(p,\theta,\varphi) = \frac{f(p)}{p}\,
\phi_{jm}(\theta,\varphi)+\frac{h(p)}{p}\,
\chi_{jm}(\theta,\varphi),
\label{ansatz}
\end{equation}
where $p$ is the radial  momentum, $\, p^2 =(p_i)^2= p_0^2 -m^2\, $ and
\begin{equation*}
 \phi_{jm} (\theta,\varphi)= \begin{pmatrix}
              \sqrt{\frac{j+m}{2j}}\,\, Y_{j-1/2}^{m-1/2}(\theta,\varphi) \\[10pt]
\sqrt{\frac{j-m}{2j}}\, \,Y_{j-1/2}^{m+1/2}(\theta,\varphi)
             \end{pmatrix} ,\qquad
\chi_{jm} (\theta,\varphi)= \begin{pmatrix}
              \sqrt{\frac{j+1-m}{2(j+1)}}\, \,Y_{j+1/2}^{m-1/2} (\theta,\varphi)\\[10pt]
-\sqrt{\frac{j+1+m}{2(j+1)}}\, \,Y_{j+1/2}^{m+1/2} (\theta,\varphi)
             \end{pmatrix} .
\end{equation*}
The $Y_l^m$ are the spherical harmonics. The $ \phi_{jm} $ and  $ \chi_{jm} $ are  orthonormal and, \cite{bjorken} 
\begin{equation}
 \begin{array}{ll}
    \phi_{jm} =\dfrac{\vec p\cdot\vec\sigma}{p} \,  \chi_{jm},\quad\ 
&  (\vec L\cdot\vec\sigma)\, \phi_{jm} = (j-\frac 12)\,\phi_{jm},
  \\[4pt]
 \chi_{jm} =\dfrac{\vec p\cdot\vec\sigma}{p} \,  \phi_{jm} ,
&
(\vec L\cdot\vec\sigma)\, \chi_{jm} = -(j+\frac 32)\,\chi_{jm}  \,  .
\end{array}
\end{equation}
 Identity $\, (\vec r\cdot\vec\sigma)(\vec p\cdot\vec\sigma)
=3i+ip\,\frac{\p}{\p p} +i\vec L\cdot\vec \sigma \, $ 
is also frequently used in the calculation.

Introducing Ansatz (\ref{ansatz}), we obtain the system
\begin{align}
& p_0 p\, \frac{df}{dp}-( j+\frac 12 )\,
 p_0 f
=\left(2i\lambda+j\right) p h  \label{1}     \\[4pt]
& p_0 p\, \frac{dh}{dp} +(j+\frac 12 )\,
p_0 h
=\left(2i\lambda - j-1\right) p f   .     \label{2}
\end{align}
Making the change of functions
\begin{equation}
  f =  p^{j+\frac 12} F\, ,\qquad
h =  p^{-j-\frac 12}  H\, ,
\end{equation}
we get the  first order system of equations
\begin{eqnarray}
&& m\, \frac{dF}{dp_0}=(2i\lambda +j)\,\left( \frac pm \right)^{-2j-2}H\, ,\\[4pt]
&& m \, \frac{dH}{dp_0}=(2i\lambda -j-1) \, \left( \frac pm \right)^{2j}F    \,  .
                                                   \label{relation}
\end{eqnarray}
The corresponding second order equations for $F$ and $H$ are
\begin{align}
& p^2 \,\frac{d^2 F}{dp_0^2} +2(j+1)p_0\,\frac{d F}{dp_0}
 -(2i\lambda+j)(2i\lambda-j-1)F =0      \, , \label{A}\\
& p^2 \,\frac{d^2 H}{dp_0^2} -2j p_0\,\frac{d H}{dp_0}
 -(2i\lambda+j)(2i\lambda-j-1) H =0  \, .         \label{B}
\end{align}

These equations can be transformed to the  Legendre equation by an additional change of functions. Introducing
\begin{equation}
x=\frac{p_0}{m}\,
\end{equation}
and
\begin{equation}
 F=(x^2-1)^{-\frac j2}\,\tilde F \, ,\qquad
 H=(x^2-1)^{ \frac{j+1}{2} }\,\tilde H\, ,
\end{equation}
we obtain
\begin{align}
& (x^2-1)\, \frac{d^2\tilde F}{dx^2} +2x\,\frac{d\tilde F}{dx}
 -\frac{j^2}{x^2-1}\,\tilde F =
2i\lambda (2i\lambda -1)\tilde F \, ,         \label{L1}\\[8pt]
& (x^2-1)\, \frac{d^2 \tilde H}{dx^2} +2x\,\frac{d\tilde H}{dx}
-\frac{(j+1)^2}{x^2-1}\,\tilde H =
2i\lambda (2i\lambda -1)\tilde H\, .            \label{L2}
\end{align}

Two linearly independent solutions to the Legendre equation (\ref{Legendre}) are the associated Legendre functions $P^\mu_\nu(x)$ and  $Q^\mu_\nu(x)$, or  $P^\mu_\nu(x)$ and  $P^{-\mu}_\nu(x)$. For (\ref{L1}-\ref{L2}), two linearly independent 
pairs of solutions are
\begin{align}
&\tilde F(x) = \tilde A \,P^j_{-2i\lambda}(x)=
(2i\lambda+j)\tilde B\,
P^j_{-2i\lambda}(x),\qquad \tilde H(x) =\tilde B\,
P^{j+1}_{-2i\lambda}(x),          \\[4pt]
&\tilde F(x) =  A\, P^{-j}_{-2i\lambda}(x)
,\qquad \tilde H(x) = B\,
P^{-j-1}_{-2i\lambda}(x) = A\, (2i\lambda -j-1)\, P^{-j-1}_{-2i\lambda}(x)\, .
       \label{solutio}
\end{align}
Relations between  coefficients $\,\tilde A$, $A$ and $\,\tilde B$, $B$ follow from (\ref{relation}) and the recurrence relations for the associated
Legendre functions. But (as we show in the Appendix) functions of the first pair diverge at  point $\,x=1\,$; therefore there is only one normalizable solution,
(\ref{solutio}),  for every real number $\,\lambda$. The corresponding radial functions $f$ and $h$ are equal to
\begin{equation}
 f_{\lambda j}=A\left( \frac p m \right)^{\frac 12}
 P^{-j}_{-2i\lambda}\left( \frac {p_0}{m} \right),
\qquad
 h_{\lambda j}=A\,(2i\lambda -j-1)
\left( \frac pm \right)^{\frac 12} P^{-j-1}_{-2i\lambda}
\left(\frac {p_0}{m} \right)   ,                     \label{solution}
\end{equation}
and they give the eigenfunctions $ \, \tilde\psi_{\lambda j m}$ of $W^4$ via (\ref{ansatz}) an (\ref{chiral}). We confirm in the Appendix that this set of eigenfunctions is complete: $ \, \tilde\psi_{\lambda j m}$ are orthogonal and normalized to $\delta$-function,
\begin{equation}
 \big(\tilde\psi_{\lambda j m},\tilde\psi_{\lambda^\prime
j^\prime m^\prime}\big) =
2A^* A^\prime \,\, \frac{\Gamma(\frac 12 -2i\lambda)\,
\Gamma(\frac 12 + 2i\lambda)} {\Gamma(j+1 -2i\lambda)\,
 \Gamma(j+1 + 2i\lambda)}\,\,
 \delta_{m m^\prime}\,
 \delta_{j j^\prime}\,\delta(\lambda -\lambda^\prime)\, ,                   \label{ON}
\end{equation}
so the normalization and the phases can be fixed as 
\begin{equation}
 A= \frac{\Gamma(j+1 + 2i\lambda)}{\sqrt 2\, \,\Gamma(\frac 12 + 2i\lambda)}\, .
\end{equation}

\section{Group-theoretic view}

In the previous section we solved the eigenvalueproblem of $W^4$ in  $(\rho, s=\frac 12)$ UIR of the principal continuous series of  $\mathrm{SO}(1,4)$ using the Hilbert space representation \cite{Moylan}. But  this problem could have been solved using the results of representation theory. Namely, the embedding coordinates,  components of the `Pauli-Lubanski' vector $W^\alpha$, coincide in fact with one of the two quadratic Casimir operators of the subgroups of $\mathrm{SO}(1,4)$: $W^0$ is a  Casimir operator of  $\mathrm{SO}(4)$ while $W^4$ and $W^i$ are Casimir operators of  $\mathrm{SO}(1,3)$ subgroups\footnote{This very important observation is due to our referee, and it gives  much better understanding of the construction of fuzzy de~Sitter space and of its structure.}. This can be easily seen from their definition:
\begin{align}
W^0 =&\,\frac 18\, \epsilon^{0\alpha\beta\gamma\delta} M_{\alpha\beta}
M_{\gamma\delta} = \frac 14\, \epsilon^{ijk}(M_{ij} M_{4k}+M_{4k}M_{ij})
\\[4pt]
W^4 =&\,\frac 18\, \epsilon^{4\alpha\beta\gamma\delta} M_{\alpha\beta}
M_{\gamma\delta} =- \frac 14\, \epsilon^{ijk}(M_{ij} M_{0k}+M_{0k}M_{ij})\, ,
\end{align}
where $\, \epsilon_{0ijk4}=\epsilon_{ijk}$. As $W^0$ is a Casimir operator of the compact group $\mathrm{SO}(4)$, it has discrete eigenvalues which are equal to $\,k'(k'+1) - k(k+1)$. On the other hand, to find the eigenvalues of $W^4$ one
has to decompose representation $(\rho,s)$ or  $(\rho,s=\frac 12)$ of the principal continuos series of  $\mathrm{SO}(1,4)$ into the UIR's of its subgroup $\mathrm{SO}(1,3)$. This was done by Str\"om, and the resulting decomposition of the representation space, $\, {\cal H}^s={\cal H}^{s+} \oplus{\cal H}^{s-}$, is in \cite{Strom:1968zz}  written as
\begin{align}
 {\cal H}^{s\pm} =& (2\pi^4)^{-2}\int\limits_0^{\infty}
\sum_{s_0=\pm s,\pm(s-1),\dots}         \nonumber
 {\cal H}^{s\pm}(s_0,\nu) \, (s_0^2+\nu^2) \, d\nu\, \\
 =&(2\pi^4)^{-2}\int\limits_{-\infty}^{\infty}
\sum_{s_0= s,s-1,\dots}
 {\cal H}^{s\pm}(s_0,\nu) \, (s_0^2+\nu^2) \, d\nu\,     \label{Sum}
\end{align}
where $s_0$ and $\nu$ label the UIR's of the Lorentz group. The representation space of the $(\rho, s)$ representation  is decomposed into a direct integral and sum  of  unitary irreducible representations $\,(\nu, s_0)$ of $\mathrm{SO}(1,3)$:  $\nu\in (-\infty, +\infty)$ is continuous  and  $s_0$, $\vert s_0\vert\leq s$, is discrete. The eigenvalue of $W^0$ which corresponds to each of the representations in decomposition (\ref{Sum}) is equal to  $\,s_0\nu$.

Our result for $s=\frac 12$ is in accordance with this. There is only one summand in (\ref{Sum}) corresponding to $\, s_0=s=\frac 12$; the spectrum of $W^0$ is the real axis,  $\, \lambda=\frac \nu 2\in(-\infty, +\infty)\, $. An analogous decomposition of  unitary irreducible representations of the Poincar\'e group into a direct integral of UIR's of the Lorentz group was done in \cite{Joos:1962qq}: as we here use the same representation space \cite{Moylan}, there are many paralells in  two calculations.

\section{Summary and outlook}

In this paper we continued our investigation of fuzzy de~Sitter space defined as a unitary irreducible representation of the de~Sitter group $\mathrm{SO}(1,4)$, analyzing representations of the principal continuous series. In analogy with the commutative case, fuzzy de~Sitter space in four dimensions is defined as an embedding in five dimensions: the embedding coordinates are proportional to components of the Pauli-Lubanski vector,  $\,x^\alpha=\ell W^\alpha$, and the embedding relation is the  Casimir relation  $\, W_\alpha W^\alpha$=\,const.  By an explicit calculation in the $\,(\rho,s=\frac 12)$ representation we found that the spectrum of time $x^0$ is discrete while the spectra of spacelike coordinates $x^4$ and $x^i$ are continuous. This result is in fact general and holds for all principal continuous UIR's $\,(\rho,s)$ of the $\,\mathrm{SO}(1,4)$, which can be proved by using the result \cite{Strom:1968zz} for the decomposition of representations of the principal series of  $\mathrm{SO}(1,4)$ into the  UIR's its $\mathrm{SO}(1,3)$ subgroup. 

There are other operators, that is other coordinates on  fuzzy de~Sitter space whose properties one would like to understand and physically interpret.
First of them is certainly the cosmological time,  $\, \tau = -\ell \log\, (W^0+W^4)$, and second are the isotropic coordinates. 
While it is, at least in the $\,(\rho,s=\frac 12)$ representation, straightforward to write the eigenvalueproblem for $\tau$, the corresponding differential  equation turns out to be not easy to solve. This is one of the problems in the given setup which  deserves additional work and which might give interesting results.

The given construction of fuzzy de~Sitter space can be straightforwardly  generalized to other spaces of maximal symmetry with the symmetry groups $\mathrm{SO}(p,q)$, in particular for even-dimensional spaces, $\, p+q=d+1\,$ with even $\,d$. In these cases,  embedding coordinates can, as for $d=4$, be  identified with the highest $W$-symbol, 
\begin{equation}
 W^\alpha = \epsilon^{\alpha \alpha_1 \alpha_2\dots
 \alpha_{d-1}\alpha_d}
M_{\alpha_1\alpha_2}\dots M_{\alpha_{d-1}\alpha_d} \, ,
\end{equation}
which is a vector in a $(d+1)$-dimensional flat space. 
The embedding relation is  the Casimir relation $\,W_\alpha W^\alpha$=\,const, and the appropriate fuzzy space is then defined as an UIR of the $\mathrm{SO}(p,q)$ group. Further, $W^\alpha$ are the Casimir operators of subgroups  $\mathrm{SO}(p-1,q)$ and  $\mathrm{SO}(p,q-1)$ and their properties are in  large part determined by the group theory. On the other hand for fuzzy Lorentzian spaces, particularly interesting are the $\mathrm{SO}(1,d)$ groups which describe  conformal symmetry in $\,d-1\,$ dimensions. Their representation theory  is well studied, in particular, the decomposition formulas for the UIR's of the principal continuous series, \cite{Boyer:1971zz} are known. Moreover, the algebra of the conformal group has the same structure that was used to define differential calculus for fuzzy de~Sitter space in four dimensions.  
Clearly, for arbitrary dimension $\, d$  momenta can be defined analogously to (\ref{444}), as generators of translations of the conformal group,
\begin{equation}
 i\tilde p_0 \sim  M_{0d}\, ,\qquad
 i\tilde p_i \sim M_{id}+M_{0i}\, ,
 \quad i=1,2,\dots d-1  \, .
\end{equation}
The $\,\tilde p_i$ mutually commute; the differential structure which
corresponds to this choice of the moving frame gives, in the commutative limit, 
 metric of the de~Siter space in $d$ dimensions. Therefore, a general construction 
with common general properties exists and should be further explored.

%

\vskip0.5cm
\begin{large}
{\bf Acknowledgement}
\end{large}
Authors are very much indebted to the referee for pointing out a mistake in the calculation of the spectrum (which was present in the first version of the paper) as well as for relating the given derivation to the decomposition of the UIR's of  $\mathrm{SO}(1,4)$ with respect to the UIR's of its subgroups.
This work was supported by the Serbian Ministry of Education, Science and Technological Development Grant ON171031, and by the COST action
MP 1405 ``Quantum structure of spacetime''.

\newpage

\begin{large}
{\bf Appendix}
\end{large}

{\bf Two additional formulas for $\mathrm{SO}(1,4)$ }

It would seem from  (\ref{[W,W]}) that operators proportional to $x^\alpha$  cannot be chosen as momenta, as it is often done in noncommutative geometry. But  it is in fact possible, in a fixed UIR, to express $M_{\alpha\beta}$ in terms of $W^\alpha$. Using  formula (which can be checked explicitly)
\begin{equation}
 W^\beta M_{\alpha\beta} = 2i W_\alpha     ,            \label{2W}
\end{equation}
we find that
\begin{equation}
i\,{\cal W}M^{\rho\sigma}=[W^\rho, W^\sigma]+
\frac12\,\epsilon^{\alpha\mu\rho\sigma\tau}W_\tau[W_\alpha,W_\mu] .
                                                          \label{3W}
\end{equation}
This also means that one can  use $W^\alpha$ as `primitive generators' \cite{Gazeau:2009mi} of fuzzy de~Sitter space.  However,  metric defined by this choice of momenta cannot be brought to the de~Sitter form, or at least it is far from obvious how to do it (a nice simple formula which expresses $M^{\alpha\delta}M^\beta{}_\delta$ in terms of $W^\alpha$ is lacking).
\\

{\bf Completeness and orthogonality of eigenfunctions $\ \tilde\psi_{\lambda j m} $}

The eigenvalue equation for $W^4$ reduces to the Legendre equation
\begin{align}
(x^2-1)\, \frac{d^2 y}{dx^2} +2x\,\frac{dy}{dx} -\frac{\mu^2}{x^2-1}\, y =
\nu(\nu +1)y                   \label{Legendre}
\end{align}
where the order $\,\mu = \pm j, \pm(j+1)\,$ is half-integer and
the degree $\,\nu = -2i\lambda\,$ is imaginary. The 
independent variable $\, x=\frac{p_0}{m}\in [1,\infty)$.

To discuss  behavior of  solutions (\ref{solutio}) at $\,x=1$ and $x=\infty$, we express the associated Legendre functions in terms of  hypergeometric function
$\, _2F_1$. For $x=1$ we use, \cite{erdelyi}
\begin{equation}
 P^\mu_\nu(z) = \frac{1}{2^\nu \Gamma(1-\mu)}\, (z+1)^{\frac \mu 2+\nu}
(z-1)^{-\frac \mu 2}\, _2F_1\left( -\nu, -\nu-\mu; 1-\mu; \,\frac{z-1}{z+1} \right)\, .
\end{equation}
As $\,_2F_1(a,b;c;0) =1$, in the vicinity of $\,x=1$  we have
\begin{equation}
  P^{\pm j}_{-2i\lambda}(x) = \frac{2^\pm \frac j2}{\Gamma(1\mp j)} \,
(x-1)^{\mp\frac j 2}\, ,
\end{equation}
that is, $P^{ j}_{-2i\lambda}\,$ is divergent and $P^{- j}_{-2i\lambda}\, $ tends to zero. Therefore we dismiss the first solution for $F$, $H$ in  (\ref{solutio}).

Similarly, at the other end of the interval $\,x=\infty\,$, we use the formula
\begin{align}
\hskip-0.5cm
P^\mu_\nu(z) = & 2^{-\nu-1} \pi^{-\frac 12}\,
\frac{\Gamma(-\frac 12-\nu)}{ \Gamma(-\nu-\mu)}\,
z^{-\nu+\mu-1}\, (z^2-1)^{-\frac \mu 2}\, _2F_1\left( \frac{1+\nu-\mu}{2},
\frac{2+\nu-\mu}{2}; \frac 32+\nu; \,\frac{1}{z^2} \right)\, \nonumber \\
& + 2^{\nu} \pi^{-\frac 12}\,
\frac{\Gamma(\frac 12+\nu)}{ \Gamma(1+\nu-\mu)}\,
z^{\nu+\mu}\, (z^2-1)^{-\frac \mu 2}\, _2F_1\left( \frac{-\nu-\mu}{2},
\frac{1-\nu-\mu}{2}; \frac 12-\nu; \,\frac{1}{z^2} \right)
\end{align}
so we have the asymptotics
\begin{equation}
 P^{-j}_\nu(x) = \frac{1}{\sqrt \pi}\, \frac{\Gamma(-\frac 12-\nu)}{\Gamma(j-\nu)}\,
(2x)^{-\nu-1} +
\frac{1}{\sqrt \pi}\, \frac{\Gamma(\frac 12+\nu)}{\Gamma(j+1+\nu)}\,
(2x)^{\nu} \, , \qquad x\to\infty \, .                  \label{xtoinfty}
\end{equation}

Let us determine the scalar product of two eigenfunctions $ \tilde\psi_{\lambda j m}$,  $\,\tilde\psi_{\lambda^\prime j^\prime m^\prime}$. They are given by (\ref{chiral}), (\ref{ansatz}),
\begin{align}
\tilde\psi_{\lambda j m}=&\begin{pmatrix}
   \tilde\chi_{\lambda j m}\\[4pt]
 \dfrac{p_0+\vec p\cdot\vec \sigma}{m}\,\tilde\chi_{\lambda j m}
 \end{pmatrix} , \qquad
\tilde \chi_{\lambda j m} = \frac{f_{\lambda j}}{p}\,
\phi_{jm}+\frac{h_{\lambda j}}{p}\,
\chi_{jm}\, .                            \label{ANS}
\end{align}
The scalar product (\ref{chiralproduct}) for Ansatz (\ref{ansatz}), (\ref{ANS}), reduces to
\begin{align}
(\tilde\psi,\tilde\psi^\prime) =&2\delta_{jj^\prime}\delta_{mm^\prime}
\int_m^\infty \frac{dp_0}{m}\,\left( {p_0}
\left(f^*\frac{df^\prime}{dp_0}+h^*\, \frac{dh^\prime}{dp_0}\right)+
\left(f^*\frac{dh^\prime}{dp_0}+h^*\, \frac{df^\prime}{dp_0}\right) \right)\, .
\end{align}
For the eigenfunctions (\ref{solution}) we find
\begin{align}
(\tilde\psi,\tilde\psi^\prime) =& 2\delta_{jj^\prime}\delta_{mm^\prime}
\int_1^\infty dx\, \Big( x\left( A^* A^\prime \, P^{-j *}_{-2i\lambda}P^{-j }_{-2i\lambda^\prime} +
B^* B^\prime  \, P^{-j -1\, *}_{-2i\lambda}P^{-j-1 }_{-2i\lambda^\prime }
 \right)
\\
& +\sqrt{x^2-1}\,\left(
 A^* B^\prime\, P^{-j *}_{-2i\lambda}P^{-j -1}_{-2i\lambda^\prime} +
B^* A^\prime \, P^{-j -1\, *}_{-2i\lambda}P^{-j}_{-2i\lambda^\prime }
\right) \Big)  \ .\nonumber
\end{align}
Using  relation
$$
\Big(\nu^\prime (\nu^\prime +1) -\nu^* (\nu^*+1) \Big)
P^{j*}_\nu P^j_{\nu^\prime} =
\frac{d}{dx}\Big( P^{j*}_\nu\,(x^2-1) \,\frac{dP^j_{\nu^\prime}}{dx}
-  P^{j}_{\nu^\prime}\,(x^2-1)\, \frac{dP^{j*}_{\nu}}{dx}
\Big)\, ,
$$
which is a consequence of the Legendre equation, and various recurrence relations between the associated Legendre functions \cite{erdelyi}, we can transform the expression under  the integral to a total derivative. We find
$$
(\tilde\psi,\tilde\psi^\prime) =\delta_{jj^\prime}\delta_{mm^\prime}\,
\frac{2A^* A^\prime\, \sqrt{x^2-1}}{2i\lambda^\prime - 2i\lambda}
\Big(-(2i\lambda+j+1)  P^{-j -1\,*}_{-2i\lambda}P^{-j }_{-2i\lambda^\prime-1}
+(2i\lambda^\prime+j+1)  P^{-j *}_{-2i\lambda}P^{-j-1 }_{-2i\lambda^\prime-1}
\Big)_1^\infty \,.
$$
This expression vanishes at the lower bound $\,x=1$, while at
the upper bound we  use the asymptotics (\ref{xtoinfty}). We obtain
\begin{align}
(\tilde\psi,\tilde\psi^\prime)  =& \delta_{jj^\prime}\delta_{mm^\prime}\,\,
\frac 1\pi\, \frac{2A^* A^\prime}{2i\lambda^\prime - 2i\lambda}
\nonumber \, \lim_{x\to\infty}\left( (2x)^{2i\lambda^\prime - 2i\lambda}\,\,
\frac{\Gamma(\frac 12 -2i\lambda)\,
\Gamma(\frac 12 + 2i\lambda^\prime)} {\Gamma(j+1 -2i\lambda)\,
\Gamma(j+1 + 2i\lambda^\prime )}\right.\, \\
& -
\left.(2x)^{-2i\lambda^\prime +2i\lambda}\,\,
\frac{\Gamma(\frac 12 +2i\lambda)\,
\Gamma(\frac 12 - 2i\lambda^\prime)} {\Gamma(j+1 +2i\lambda)\,
\Gamma(j+1 - 2i\lambda^\prime )}
\right)  \, .  \nonumber
\end{align}
For $\,a =2\lambda^\prime-2 \lambda\neq 0\,$ the upper expression, as a distribution that is `under the integral', vanishes
\begin{equation}
 \lim\limits_{x\to\infty} x^{ia} =0, \qquad a\neq 0 \, .
\end{equation}
For $a\to 0\,$ we can use 
\begin{equation}
  \lim\limits_{x\to\infty} \frac{x^{ia}-x^{-ia}}{2\pi i a} =\delta(a)
\end{equation}
to obtain normalization condition (\ref{ON}).

\end{document}